\documentclass[twocolumn]{aastex631}
\usepackage{amsmath, amstext, verbatim, graphicx, xcolor, longtable}
\usepackage{float}
\usepackage{comment}

\newcommand{\Hb}{\hbox{{\rm H}$\beta$}}
\newcommand{\Ha}{\hbox{{\rm H}$\alpha$}}
\newcommand{\NII}{\hbox{{\rm [N}\kern 0.1em{\sc ii}{\rm ]}}}
\newcommand{\OIII}{\hbox{{\rm [O}\kern 0.1em{\sc iii}{\rm ]}}}

\begin{document}
\title{\large \bf Beyond the Monsters: A More Complete Census of Black Hole Activity \\ at Cosmic Dawn}
\shorttitle{Intrinsic BH Populations}
\shortauthors{Brooks et al.}

%% author blocks 

\author[0000-0001-5384-3616]{Madisyn Brooks}
\altaffiliation{NSF Graduate Research Fellow}
\affil{Department of Physics, 196A Auditorium Road, Unit 3046, University of Connecticut, Storrs, CT 06269, USA}

\author[0000-0002-1410-0470]{Jonathan R. Trump}
\affil{Department of Physics, 196A Auditorium Road, Unit 3046, University of Connecticut, Storrs, CT 06269, USA}

\author[0000-0002-6386-7299]{Raymond C.\ Simons}
\affiliation{Department of Engineering and Physics, Providence College, 1 Cunningham Sq, Providence, RI 02918 USA}

\author[0000-0002-6348-1900]{Justin Cole}
\altaffiliation{NASA FINESST Investigator}
\affiliation{Department of Physics and Astronomy, Texas A\&M University, College Station, TX 77843-4242, USA}
\affiliation{George P. and Cynthia Woods Mitchell Institute for Fundamental Physics and Astronomy, Texas A\&M University, College Station, TX 77843-4242, USA}

\author[0000-0003-1282-7454]{Anthony J. Taylor}
%\email{anthony.taylor@austin.utexas.edu}
\affiliation{Department of Astronomy, The University of Texas at Austin, Austin, TX, USA}
\affiliation{Cosmic Frontier Center, The University of Texas at Austin, Austin, TX, USA}

\author[0000-0002-9921-9218]{Micaela B. Bagley}
\affiliation{Department of Astronomy, The University of Texas at Austin, Austin, TX, USA}
\affiliation{Astrophysics Science Division, NASA Goddard Space Flight Center, 8800 Greenbelt Rd, Greenbelt, MD 20771, USA}

\author[0000-0001-8519-1130]{Steven L. Finkelstein}
\affiliation{Department of Astronomy, The University of Texas at Austin, Austin, TX, USA}
\affiliation{Cosmic Frontier Center, The University of Texas at Austin, Austin, TX, USA}

\author[0000-0001-8047-8351]{Kelcey Davis}
\altaffiliation{NSF Graduate Research Fellow}
\affiliation{Department of Physics, 196A Auditorium Road, Unit 3046, University of Connecticut, Storrs, CT 06269, USA}
\affil{Los Alamos National Laboratory, Los Alamos, NM 87545, USA}

%% alpha order authors 
%% please add your author blocks!

\author[0000-0001-5758-1000]{Ricardo O. Amor\'{i}n} 
\affiliation{Instituto de Astrof\'{i}sica de Andaluc\'{i}a (CSIC), Apartado 3004, 18080 Granada, Spain}

\author[0000-0001-8534-7502]{Bren E. Backhaus}
\affil{Department of Physics and Astronomy, University of Kansas, Lawrence, KS 66045, USA}

\author[0000-0001-7151-009X]{Nikko J. Cleri}
\affiliation{Department of Astronomy and Astrophysics, The Pennsylvania State University, University Park, PA 16802, USA}
\affiliation{Institute for Computational and Data Sciences, The Pennsylvania State University, University Park, PA 16802, USA}
\affiliation{Institute for Gravitation and the Cosmos, The Pennsylvania State University, University Park, PA 16802, USA}
% \email{cleri@psu.edu}

\author[0000-0002-7831-8751]{Mauro Giavalisco}
\affiliation{University of Massachusetts Amherst, 710 North Pleasant Street, Amherst, MA 01003-9305, USA}

\author[0000-0001-9440-8872]{Norman A. Grogin}
\affiliation{Space Telescope Science Institute, 3700 San Martin Drive, Baltimore, MD 21218, USA}

\author[0000-0002-3301-3321]{Michaela Hirschmann}
\affiliation{Institute of Physics, Laboratory of Galaxy Evolution, Ecole Polytechnique Fédérale de Lausanne (EPFL), Observatoire de Sauverny, 1290 Versoix, Switzerland}

\author[0000-0002-4884-6756]{Benne W. Holwerda}
\affiliation{Department of Physics, University of Louisville, Natural Science Building 102, 40292 KY Louisville, USA}

\author[0000-0002-1416-8483]{Marc Huertas-Company}
\affiliation{Instituto de Astrofísica de Canarias (IAC), La Laguna, E-38205,
Spain}
\affiliation{Universidad de La Laguna. Avda. Astrofísico Fco. Sanchez, La La-
guna, Tenerife, Spain}
\affiliation{Observatoire de Paris, LERMA, PSL University, 61 avenue de
l’Observatoire, F-75014 Paris, France}
\affiliation{Université Paris-Cité, 5 Rue Thomas Mann, 75014 Paris, France}

\author[0000-0001-9187-3605]{Jeyhan S. Kartaltepe}
\affiliation{Laboratory for Multiwavelength Astrophysics, School of Physics and Astronomy, Rochester Institute of Technology, 84 Lomb Memorial Drive, Rochester, NY 14623, USA}

\author[0000-0002-8360-3880]{Dale D. Kocevski}
\affiliation{Department of Physics and Astronomy, Colby College, Waterville, ME 04901, USA}

\author[0000-0002-6610-2048]{Anton M. Koekemoer}
\affiliation{Space Telescope Science Institute, 3700 San Martin Drive,
Baltimore, MD 21218, USA}
%\email{koekemoer@stsci.edu)

\author[0000-0003-1581-7825]{Ray A. Lucas}
\affiliation{Space Telescope Science Institute, 3700 San Martin Drive, Baltimore, MD 21218, USA}

\author[0000-0001-9879-7780]{Fabio Pacucci}
\affiliation{Center for Astrophysics $\vert$ Harvard \& Smithsonian, 60 Garden St, Cambridge, MA 02138, USA}
\affiliation{Black Hole Initiative, Harvard University, 20 Garden St, Cambridge, MA 02138, USA}
%\email[show]{fabio.pacucci@cfa.harvard.edu}

\author[0000-0002-9373-3865]{Xin Wang}
\affiliation{School of Astronomy and Space Science, University of Chinese Academy of Sciences (UCAS), Beijing 100049, China}
\affiliation{National Astronomical Observatories, Chinese Academy of Sciences, Beijing 100101, China}
\affiliation{Institute for Frontiers in Astronomy and Astrophysics, Beijing Normal University, Beijing 102206, China}

\begin{abstract}
JWST has revealed an abundance of low-luminosity active galactic nuclei (AGN) at high redshifts ($z > 3$), pushing the limits of black hole (BH) science in the early Universe. Results have claimed that these BHs are significantly more massive than expected from the BH mass-host galaxy stellar mass relation derived from the local Universe. We present a comprehensive census of the BH populations in the early Universe through a detailed stacking analysis of galaxy populations, binned by luminosity and redshift, using JWST spectroscopy from the CEERS, JADES, RUBIES, and GLASS extragalactic deep field surveys. Broad \Ha\ detections in $31\%$ of the stacked spectra (5/16 bins) imply median BH masses of $10^{5.21} - 10^{6.13}~ \rm{M_{\odot}}$ and the stacked SEDs of these bins indicate median stellar masses of $10^{7.84} - 10^{8.56} ~\rm{M_{\odot}}$. This suggests that the median galaxy hosts a BH that is \textit{at most} a factor of 10 times over-massive compared to its host galaxy and lies closer to the locally derived $M_{BH}-M_*$ relation. We investigate the seeding properties of the inferred BHs and find that they can be well-explained by a light stellar remnant seed undergoing moderate Eddington accretion. Our results indicate that individual detections of AGN are more likely to sample the upper envelope of the $M_{BH}-M_*$ distribution, while stacking on ``normal" galaxies and searching for AGN signatures can overcome the selection bias of individual detections.

\end{abstract}

\section{Introduction}\label{Introduction}
The routine detection of quasars in both the local and high redshift ($z > 3$) Universe has provided strong evidence for the presence of a black hole (BH) in the center of almost every massive galaxy. Measurements of galactic properties, like velocity dispersion and bulge mass, have shown that these BHs likely play a critical role in their host galaxy's evolution across cosmic time \citep[e.g.,][]{Magorrian1998,KormendyHo2013,ReinesVolonteri2015, Greene2020}. More recent studies with JWST have pushed the boundary of BH science by detecting active galactic nuclei (AGN) in abundance through their broad Balmer emission lines \citep[e.g.,][]{Kocevski2023, Larson2023, Harikane2023, Maiolino2023, Ubler2023, Killi2023, Kokorev2023, Greene2023, Kocevski2024, Taylor2024, Naidu2025, Taylor2025, Hviding2025}.

In stark contrast to the tight relationship between black hole mass ($\rm{M_{BH}}$) and its host galaxy's stellar mass ($\rm{M_*}$) in the local Universe \citep[e.g.,][]{ReinesVolonteri2015}, results from JWST have shown that most of these AGN are $10-100$ times more massive than expected compared to their galactic host’s stellar mass \citep[e.g.,][]{Maiolino2023, Pacucci2023, Harikane2023, Ubler2023, Larson2023, Durodola2025, Chen2025, Jones2025}, but some still agree with the stellar velocity dispersion relation \citep{Maiolino2023, Cohn2025, Shankar2025}.  The detection of these ``over-massive" BHs (offset from the local $M_{BH}-M_*$ relation) could be expected with a careful consideration of selection bias; only the most luminous AGN can be observed with flux-limited surveys \citep{Lauer2007}, and will tend to be over-massive compared to the local relation \citep{Li2025, Sun2024, Shankar2025}. However, detailed studies show that these early Universe BHs could be fundamentally different from the low-redshift quasars that structure the local $M_{BH}-M_*$ relation\citep[e.g.,][]{Wyithe2003, Caplar2018, Yang2018, Inayoshi2024, Pacucci2025}. \cite{Pacucci2023} found that selection bias can account for \textit{at most} 0.2~dex of the offset above the local relation and \cite{Stone2024} found that $z\sim6$ quasars are statistically distinct from very luminous local AGN, which also suffer from selection bias effects.

 The progenitors of these BHs are categorized into light ($\lesssim 10^{3}~\rm{M_{\odot}}$) seeds, which result from Pop III stellar remnants, and heavy ($\gtrsim 10^{4}~\rm{M_{\odot}}$) seeds, which formed from the direct collapse of metal-free gas clouds or through runaway collisions in dense stellar environments \citep[e.g.,][]{Volonteri2003,BrommLoeb2003,Lodato2006,Inayoshi2020, Klessen2023}. The formation of BHs through the classic light seeding channel, with steady Eddington-limited accretion, is insufficient to explain the masses of the observed JWST AGN \citep[e.g.,][]{Larson2023,Pacucci2023, Taylor2024, Taylor2025}.  If the high-redshift JWST AGN are truly over-massive compared to the derived local relation, these sources would provide strong evidence for heavy BH seeds \citep{Larson2023, Natarajan2024} or the growth of light seeds through super-Eddington accretion \citep{Lambrides2024, Pacucci2024, Taylor2025, Jeon2025}. 
 
 Determining if typical high-redshift AGN are truly over-massive compared to their host galaxy requires probing the full range of observable BHs. Detecting BHs at these high redshifts minimizes the time between BH formation and observation, allowing us to further study the properties of the early BH seeds.Constraining the $M_{BH}-M_*$ relation in the early Universe will answer three prominent astrophysical questions: (i) are typical high-redshift AGN truly over-massive compared to their host galaxy, (ii) How did the first BHs form (see \cite{Scoggins2024}), and (iii) what shape does BH-galaxy co-evolution take in the early Universe?

 This work leverages the large sample of NIRSpec grating spectroscopy across the CEERS, JADES, RUBIES, and GLASS extragalactic deep field surveys. We construct a sample of $\sim 2000$ galaxies to search for broad \Ha\ emission lines in the undetected broad line (BL) population to probe the low-mass range of observable BHs in the high-redshift Universe. We search for low-mass BHs too faint for detection in individual spectra with a robust stacking analysis and present 5 detections of median BH masses with $ M_{BH} =10^{5.21} - 10^{6.13}~ \rm{M_{\odot}}$. We also derive BH mass upper limits for the stacks with no detected broad \Ha\ emission.

 \par This paper is presented as follows. In \S \ref{Observational Dataset} we describe our observational dataset, in \S \ref{sec: stacked spectra} we discuss our methodology for stacking spectra, in \S \ref{sec: Emission Line Fitting} we describe our emission line fitting technique, in \S \ref{sec: results} we discuss our BH mass results, and in \S \ref{sec: discussion} we describe the implications of this work. For this study, we assume a flat $\mathrm{\Lambda}$CDM cosmology with $\mathrm{H_{0}}$ = 67.4 km s $^{-1}$ Mpc$^{-1}$ and $\Omega_{\mathrm{M}} = 0.315$ \citep{Planck2020}.

\section{Observational Dataset}\label{Observational Dataset}

\begin{figure}
    \centering
    \includegraphics[width=\linewidth]{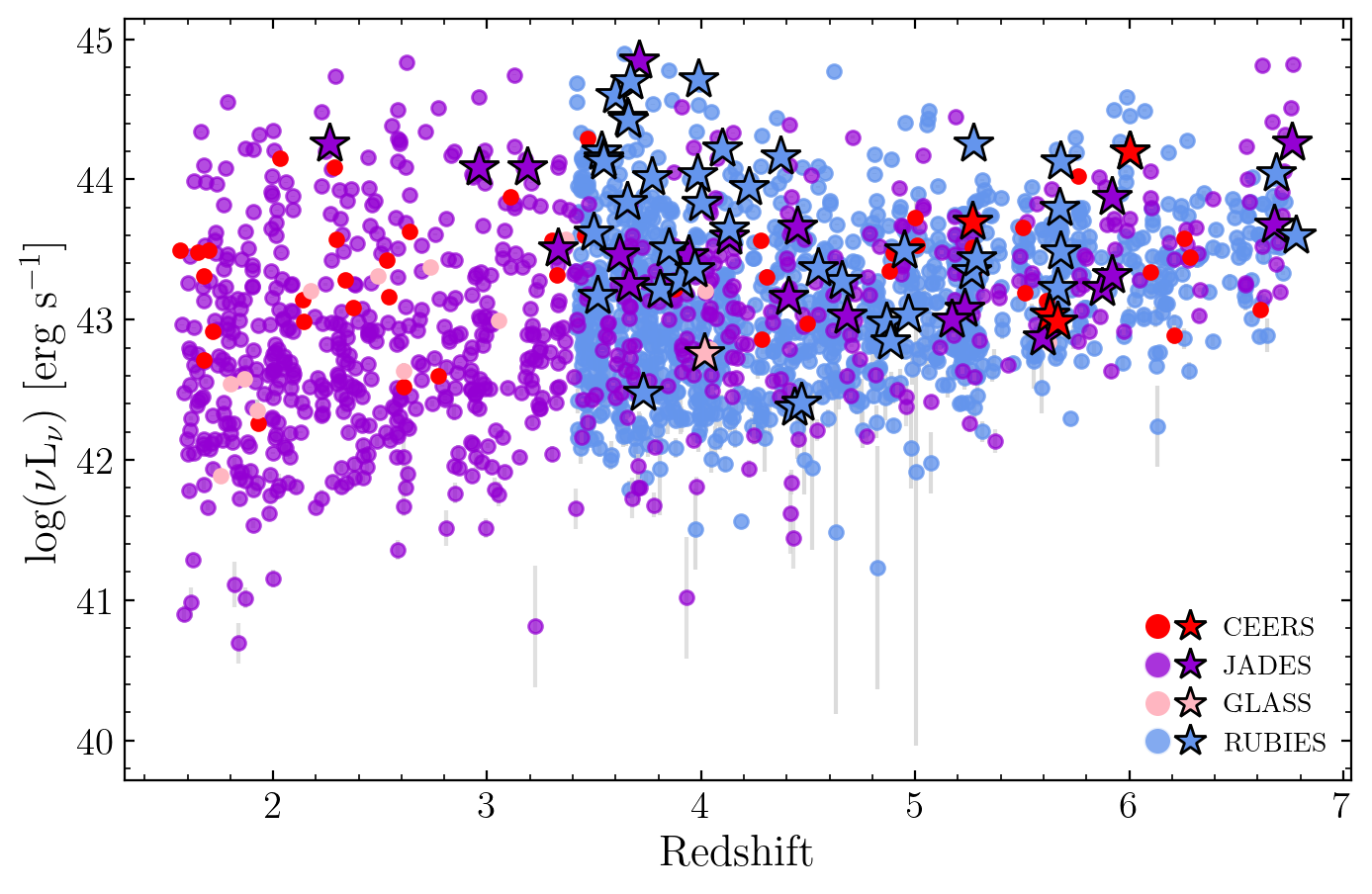}
    \caption{Continuum luminosity ($\nu L_{\nu}$) as a function of redshift for our complete galaxy sample ($\sim 2000$ galaxies). Continuum luminosities are measured from the reddest emission-line free NIRCam filter available for each source. CEERS sources are shown in red, JADES sources are shown in purple, GLASS sources are shown in pink, and RUBIES sources are shown in light blue. BL AGN identified in our sample are shown as stars and are removed from our stacking analysis.}\label{fig:sample}
\end{figure}

The analysis in this paper utilizes both \textit{JWST}/NIRCam photometry and \textit{JWST}/NIRSpec spectroscopy. Our sample consists of galaxies observed with NIRSpec medium resolution ($\lambda / \Delta \lambda$ $\sim$1000) and high resolution ($\lambda / \Delta \lambda$ $\sim$2700) spectra across the $z = 1.56 - 6.77$ redshift range. We use NIRSpec multi-object spectroscopy taken by four surveys: ERS 1345 (CEERS, PI: Steven Finkelstein), ERS 1324 (GLASS, PI: Tommaso Treu), GTO 1180, 1181, 1210, 1286 (JADES, PIs: Daniel Eisenstein, N. Lutzgendorf), and GO 4233 (RUBIES, PIs: Anna de Graaff, Gabriel Brammer). We additionally use NIRCam photometry from: ERS 1345 (CEERS, PI: Steven Finkelstein), GO 2561 (UNCOVER, PIs: Ivo Labbe, Rachel Bezanson), GTO 1180, 1181, 1210, 1286 (JADES, PIs: Daniel Eisenstein, N. Lutzgendorf), and GO 1837 (PRIMER, PI: James Dunlop). 

We identify broad-line (BL) AGN from our final sample (and remove them from our stacks) based on the sources identified in \cite{Harikane2023}, \cite{Kocevski2023}, \cite{Maiolino2023}, \cite{Taylor2024}, and \cite{Juodzbalis2025}. In total, this represents a sample of 66 BL AGN. Their AGN selection criteria are based on a dual narrow+broad Gaussian model to successfully reproduce the \Ha\ emission line, while the \OIII$\lambda5007$ emission line is reproduced by a narrow-line (NL) only model. We refer to the respective papers for a full description of their BL AGN identification process. We include only the BL AGN that have NIRCam photometry, which generally represents only a subset of BL AGN identified in previous work. %%We note that the full BL AGN samples presented in these papers might not be represented in this analysis due to insufficient photometric overlap. 
Our complete spectroscopic sample is shown in Figure \ref{fig:sample}. A brief description of the observation programs used in this study follows.

\subsection{Spectroscopy}

\subsubsection{CEERS}

The Cosmic Evolution Early Release Science Survey (CEERS) \citep{Finkelstein2025} covered $\sim$100 $\rm{arcmin^2}$ of the Extended Groth Strip (EGS) \citep{CEERSdoi} and observed with six NIRSpec pointings. The pointings were observed with the G140M/F100LP, G235M/F170LP, and G395M/F290LP grating/filter pairs, and for 0.86~hr per grating. Sources selected from the CEERS spectroscopy have robust redshifts which rely on multiple emission line detections. The spectroscopic data were processed with the STScI JWST Calibration Pipeline\footnote{\url{https://github.com/spacetelescope/jwst}} version v1.8.5 \citep{Bushouse2022}. We refer to \cite{ArrabalHaro2023} for a full description of the NIRSpec data reduction for CEERS.

\subsubsection{JADES}
The JWST Advanced Deep Extragalactic Survey (JADES) covered  $\sim$175~$\rm{arcmin^2}$ in the GOODS-S and GOODS-N fields \citep{Eisenstein2023, JADESdoi}. We use NIRSpec spectroscopy observed with the G140M/F070LP, G235M/F170LP, and G395M/F290LP grating/filter pairs, released publicly as part of the JADES DR3 \citep{D'Eugenio2025}. The spectroscopic data were processed with a custom pipeline described in \cite{Ferruit2022}. We select sources from the publicly available JADES DR3 strong emission-line flux catalogs that have well-detected redshifts; we use sources with the flags ``A,'' which have at least one emission line in the medium-resolution grating, and ``B,'' which have two or more prism emission lines. The full redshift vetting process for the JADES sources is described in \cite{D'Eugenio2025}. 

The JADES observations were split into three visits, with individual objects being observed with either one, two, or three visits at 2.3~hr exposure time per visit. Objects observed in each of the three visits reached up to $\sim$7~hr of exposure time. We refer to \cite{Eisenstein2023}, \cite{Bunker2023}, and \cite{D'Eugenio2025}
for a full description of the JADES survey and their NIRSpec
data reduction process.

\subsubsection{RUBIES}
The Red Unknowns: Bright Infrared Extragalactic Survey (RUBIES) observed 6 NIRSpec pointings in the CEERS (EGS) field and 6 NIRSpec pointings in the PRIMER-UDS field \citep{deGraaff2025}. We use the G395M/F290LP
grating/filter pair from RUBIES and each pointing had an exposure time of 0.8~hr. We select sources with redshifts published on the DJA that have a redshift grade of ``3," which is described as a robust redshift in \cite{deGraaff2025}.

The spectroscopic data were processed with the STScI JWST Calibration Pipeline version 1.13.4. We use the reduced 1D spectra produced by \cite{Taylor2024} and refer to that paper for a full description of the NIRSpec data reduction.

\subsubsection{GLASS}

The Grism Lens Amplified Survey from Space (GLASS) observed the Abell 2744 Hubble Frontier Lensing Cluster \citep{GLASSsurvey}. GLASS spectroscopy used in this work consists of one NIRSpec pointing taken with the high-resolution grating/filter pairs:  G140H/F100LP, G235H/F170LP, and G395M/F290LP. For each grating/filter pair, the exposure time was 4.9~hr. We select sources from the publicly available GLASS emission line detection catalog that have well-detected redshifts; we use sources with the flags ``4," which have multiple emission lines, and ``3," which have a single unambiguous spectral emission feature. The full redshift vetting process for the GLASS spectroscopy is described in \cite{Mascia2024}.  

We use GLASS NIRSpec data publicly available through the DAWN JWST Archive (DJA) \citep{DJA}. The spectroscopic data were processed with a custom pipeline \texttt{MsaExp} v.0.6.7 and we refer to \cite{Heintz2024, Heintz2025} and \cite{deGraaff2025} for a full description of the NIRSpec data reduction.

\subsection{Photometry}
For this study, we use NIRCam photometry from the CEERS, JADES, PRIMER (PI: James Dunlap), and UNCOVER (PIs: Ivo Labbe, Rachel Bezanson) \citep{UNCOVER} surveys to estimate continuum luminosities and to construct our stacked SEDs (described in \S \ref{sec: stacked spectra}. Additionally, we include HST imaging in the F814W band using publicly available data from the CANDELS survey \citep{Grogin2011,Koekemoer2011}. We use 7 NIRCam filters that were observed in all four surveys: F115W, F150W, F200W, F277W, F356W, F410M, and F444W.

The photometry in CEERS, JADES, and PRIMER was measured by the procedure described in \cite{Bagley2023} and \cite{Finkelstein2024}. We also use the publicly available UNCOVER DR3 photometric catalog to provide NIRCam coverage of our GLASS sources. We refer to \cite{Seuss2024} and \cite{UNCOVER} for a full description of the UNCOVER photometric data reduction.

%%\begin{deluxetable*}{cccc}
%%\label{Observational Dataset -- Photometry} 
%%\tablecaption{Observational Dataset}
%%\tablehead{\colhead{Survey} & \colhead{Field} & %%\colhead{NIRCam Filters} &\colhead{N}}
%%
%%\startdata
%%CEERS &  EGS &  F115W, F150W, F200W, F227W, F356W, F410M, F444W & \\
%%JADES & GDS, GDN  & F090W, F115W, F150W, F200W, F227W, F335M, F356W, F410M, F444W & \\
%%UNCOVER & Abell 2744 &  F115W, F150W, F200W, F277W, F356W, F410M, F444W & \\
%%PRIMER & UDS & & \\
%%\enddata
%%
%%\end{deluxetable*}

\section{Stacked Spectra}\label{sec: stacked spectra}

\begin{figure*}
    \centering
    \includegraphics[width=\linewidth]{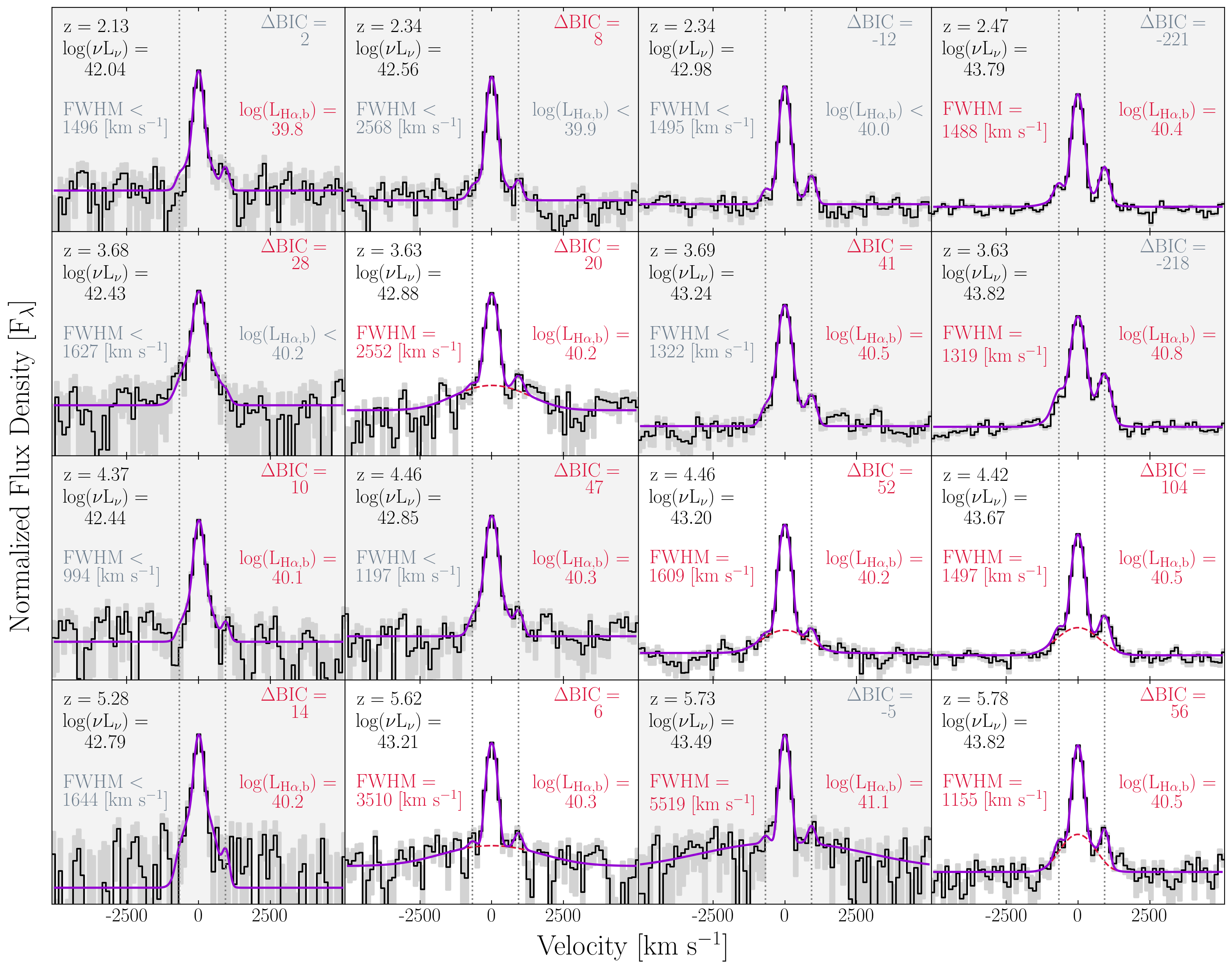}
    \caption{Median \Ha\ emission lines for our 16 stacks. Stacks increase in continuum luminosity from left to right, and increase in redshift from top to bottom. The total best-fit model is shown in purple for each stack, and a well-detected ($>3\sigma$) BL fit component is shown by the dashed red line. The $\rm{\Delta BIC}$, FWHM, and $\rm{L_{H\alpha, broad}}$ is given for each stack. Stacks that pass our three criteria for a detected BL, described in \S \ref{sec: Emission Line Fitting}, have their $\rm{\Delta BIC}$, FWHM, and $\rm{L_{H\alpha, broad}}$ shown in red and have a white background. Stacks that do not pass the BL identification criteria have a light gray background. }
    \label{fig:halpha stacks}
\end{figure*}

\begin{figure*}
    \centering
    \includegraphics[width=\linewidth]{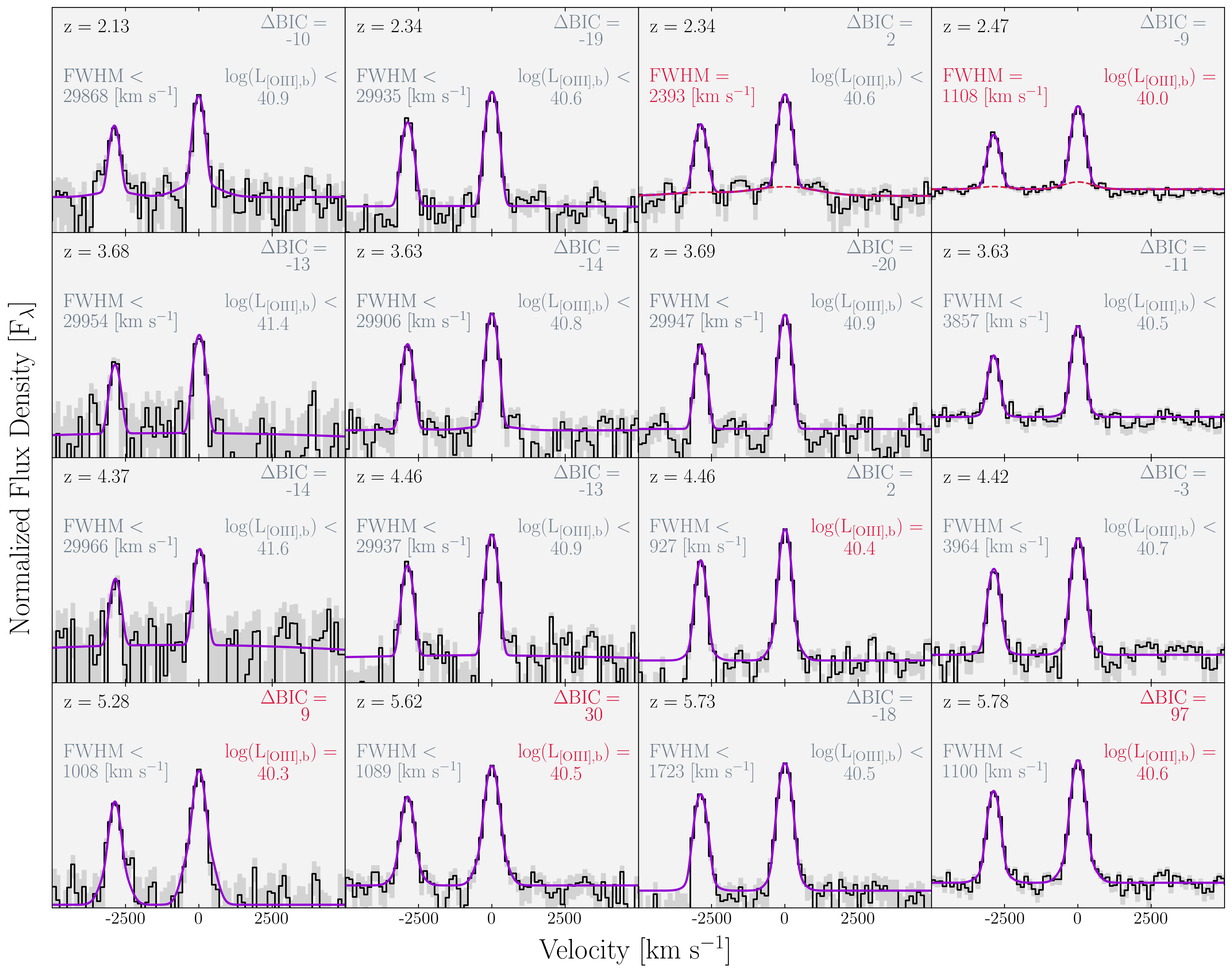}
    \caption{Median \OIII\ emission lines for our 16 stacks. Stacks increase in continuum luminosity from left to right, and increase in redshift from top to bottom. The total best-fit model is shown in purple for each stack, and a well-detected ($>3\sigma$) BL fit component is shown by the dashed red line.  The $\rm{\Delta BIC}$, FWHM, and $\rm{L_{H\alpha, broad}}$ is given for each stack. Stacks that pass our three criteria for a detected BL, described in \S \ref{sec: Emission Line Fitting}, have their $\rm{\Delta BIC}$, FWHM, and $\rm{L_{H\alpha, broad}}$ shown in red. The lack of BL \Ha\ detections indicates evidence for no outflows in these stacks. Stacks that do not pass the BL identification criteria have a light gray background.}
    \label{fig:[OIII] stacks}
\end{figure*}

We produce median-stacked NIRSpec spectra of the \Ha\ and \OIII\ emission line region in 16 redshift and continuum luminosity bins, with an equal number of sources in each bin ($\sim 125$ galaxies). We calculate continuum luminosities from the reddest emission-line-free NIRCam filter available for each source. In brief, we choose a NIRCam filter redder than \Ha\ if available. If not, we pick a NIRCam filter without \Ha\ , \Hb\ , or \OIII\ emission. To create our 16 stacks, we first create 4 redshift bins. We then slice each redshift bin into 4 continuum luminosity bins based on luminosity percentiles of the sub-sample, for a total of 16 bins. The median redshift and median continuum luminosity for each bin are shown in Table \ref{table: measurements}.

We measure initial redshifts for sources with a well-detected \Ha\ emission line ($>3\sigma$) using a Levenberg-Marquardt least-squares method implemented through the \texttt{astropy} Gaussian modeling routine \citep{Astropy2013, Astropy2018, Astropy2022}. If the source does not have a well-detected \Ha\ emission line, we rely on reported redshifts from the respective publicly available spectroscopic catalogs. To stack our sources, we interpolate the spectra to a common velocity grid ($\sim 127 \rm{~km~s^{-1}}$ per pixel), where the resolution is determined by the lowest redshift ($z = 1.58$) source of our sample. Each spectrum is then normalized by the peak of the \Ha\ line flux. We then combine the spectra by taking the median flux at each velocity-pixel position in the velocity grid. We additionally apply a slit loss/path-loss correction by applying a flux-correction factor of 1.28 to our final stacks, which is the median corrective factor found in \cite{Taylor2024}. We calculate the error of the stacked spectrum by dividing each median-stacked error pixel by $\sqrt{N}$, where N is the number of sources in each stack ($\sim 125$). Our final NIRSpec \Ha\ median stacks are shown in Figure \ref{fig:halpha stacks}, along with their best-fit Gaussian model that is described in \S \ref{sec: Emission Line Fitting}. 

We also note that spectral stacking an emission line near the edge of the filter bandwidth can lead to issues with continuum measurements and this issue is commonly seen in H\,\textsc{i} stacking (Featherstone et al., in prep; Hoosain et al., in prep; \cite{Healy2019}). The \Ha\ line is observable in the 395M NIRSpec filter up to $z =6.77$. Our highest redshift stack ($z_{median} = 5.78$) has two sources with $z > 6.75$. Both of these sources have sufficient ($\pm 0.05~ \mu m$) spectral coverage around the \Ha\ emission line, so we elect to keep these sources in our stack. 

We also produce median-stacked spectral energy distributions (SEDs) from the NIRCam photometry and the HST/WFC3 F814W band of our sources to estimate median stellar masses for our sample. We use the 7 overlapping filters across all four surveys: F115W, F150W, F200W, F227W, F356W, F410M, and F444W with the addition of the HST F814W filter. We first convert each source's SED to rest-frame by dividing both the flux density ($F_\nu$) and filter wavelengths by $(1+z)$. We normalize the fluxes in each filter to the F814W filter before median-stacking on a common rest-frame wavelength grid. After stacking, we re-normalize the SEDs by the median F814W flux. The errors for the stacked SEDs are estimated in the same manner as the NIRSpec stacks.

\subsection{Stellar Masses}\label{sec: stellar masses} We use the SED-fitting code \texttt{bagpipes} \citep{Carnall2018} to estimate the median stellar mass of each NIRCam-stacked SED, fixing the redshift of each stack to the median redshift of the galaxies in the stack (see Table \ref{table: measurements} for the $z_{median}$). For our SED fits, we assume the stacks are dominated by starlight and do not include an AGN component. BAGPIPES utilizes a $\chi^2$-likelihood function, assuming that photometric errors follow a Normal distribution, and allows for a wide range in the prior parameter spaces. BAGPIPES then samples the posteriors for model parameters using the \textit{Nautilus} sampling algorithm.  %(see "Probably Adam's most recent paper or the Nautilus paper?").

We fit all galaxies assuming a \cite{Chabrier2003} IMF and we allow BAGPIPES to fit for $\log(M_\star/M_\odot) \in [6,15]$ with a uniform prior, providing a wide prior space for stellar-mass measurements, and we assume a \cite{Calzetti2000} dust law. We used binary population and spectral synthesis (BPASS, v2.2.1; \cite{Eldridge2017}) stellar templates and fit over a range of metallicity $(Z/Z_\odot) \in [0.001, 1]$. We further allow for nebular emission (calculations were performed with version C17.vr of Cloudy, last described by \cite{Ferland2017}), where the metallicity of the nebular gas is equal to that of the stellar population ($Z_{\rm stars} = Z_{\rm neb}$). The ionization parameter ranges from $\log(U) \in [-4,-1]$ with a log-uniform prior.

For the SFH parameterization in BAGPIPES, we adopt the Gaussian process (GP) model from Dense Basis \citep{Iyer2019...879..116I}. For a more detailed description of the GP implementation, see \cite{Carnall2018}. In this case, the SFH is defined by $N$ parameters, \{$t_X$\}, where $X = 1,K,N$. Each \{$t_X$\} is the lookback time at which the fraction, $1/N$, of the total stellar mass is formed. We adopt $N = 4$ and fit for \{$t_X$\} such that each lookback time bin contains a quarter (25\%) of the total stellar mass formed, and we set a Dirichlet continuity prior (see \citealt{Leja2017, Leja2019, Iyer2019...879..116I} for a more detailed description) on the SFH, forcing continuity across the time bins. Our stacks and best-fit SEDS are shown in Figure \ref{fig:stellar mass SEDs} in the Appendix.

\section{Emission Line Fitting}\label{sec: Emission Line Fitting}
We fit the \Ha\ emission line in each stack with a dual-component, narrow+broad, Gaussian model using the Markov Chain Monte Carlo (MCMC) routine from the Python \texttt{emcee} package \citep{emcee}. The dual-component Gaussian model, adopting the methodology used in \cite{Larson2023}, is:

\begin{equation}
\begin{split}
    f(\lambda)_{dual} = f_{c} + f_{nar} \exp{\bigg(-\frac{1}{2}\frac{(\lambda - \lambda_{0})}{\sigma^{2}_{nar}}}\bigg) \\
    + f_{broad} \exp{\bigg(-\frac{1}{2}\frac{(\lambda - \lambda_{0})}{\sigma^{2}_{broad}}}\bigg)
\end{split}
\end{equation}

\noindent where $f_{c}$ is the continuum flux, which is assumed to be constant, $f_{nar}$ and $f_{broad}$ are the narrow and broad-line amplitudes, and $\sigma^{2}_{nar}$ and $\sigma^{2}_{broad}$ are the individual line widths. We model the \NII$\lambda\lambda$6550, 6585 doublet around the \Ha\ line as two Gaussians with a fixed line flux ratio of 1:2.94 \citep{Osterbrock2006}. The line widths and line centers of the \NII\ doublet are fixed by the narrow \Ha\ line. 

We run \texttt{emcee} using 16 walkers and 20,000 steps and implement a ``burn-in" of 10,000 steps that are discarded from our final analysis. In our model, we implement flat priors for each of our parameters and use the following constraints on our priors:
\begin{enumerate}
    \item $f_{nar} > 0$ and $f_{broad} > 0$
    \item $f_{nar} > f_{broad}$
    \item $\rm{FWHM}_\textit{{nar}} < 700$ $\rm{km}$ $\rm{s^{-1}}$
    \item $ 700 ~\rm{km ~s^{-1}}<\rm{FWHM}_\textit{{broad}} < 10,000$ $\rm{km}$ $\rm{s^{-1}}$
    
\end{enumerate}
We then allow $\lambda$ to vary within $500~\rm{km~s^{-1}}$ of the previously fitted line center.

To test the significance of the BL component in our \Ha\ emission line fits, we re-fit each stack with a modified narrow-line only version of our dual-component Gaussian model. We then compute the Bayesian Information Criterion (BIC) for both \Ha\ emission line models, with and without a BL component. The BIC allows us to compare the goodness-of-fit between our two models while including differing degrees of freedom. The BIC is defined as:
\begin{equation}
    BIC = \chi^2 + k \ln(N)
\end{equation}
where $k$ is the number of free model parameters and $N$ is the number of spectral pixels (data points) included in the fit. To accept the dual-Gaussian model as our favored model, we require a $3\sigma$ detection of the broad flux component, a BL FWHM of $> 1000 \rm{~km~s^{-1}}$, and a $\Delta BIC > 6$. A $\Delta BIC > 6$ provides ``strong" evidence for the inclusion of the broad-component to our \Ha\ emission line model \citep{Liddle2004}. $\Delta BIC$ is defined here as $BIC_{narrow} - BIC_{broad+narrow}$. Using these three criteria, we find significant evidence for a broad \Ha\ component in 4/16 of our median stacks, indicated by the white panels in Figure \ref{fig:halpha stacks}.

\begin{figure*}[ht!]
    \centering
    \includegraphics[width=\linewidth]{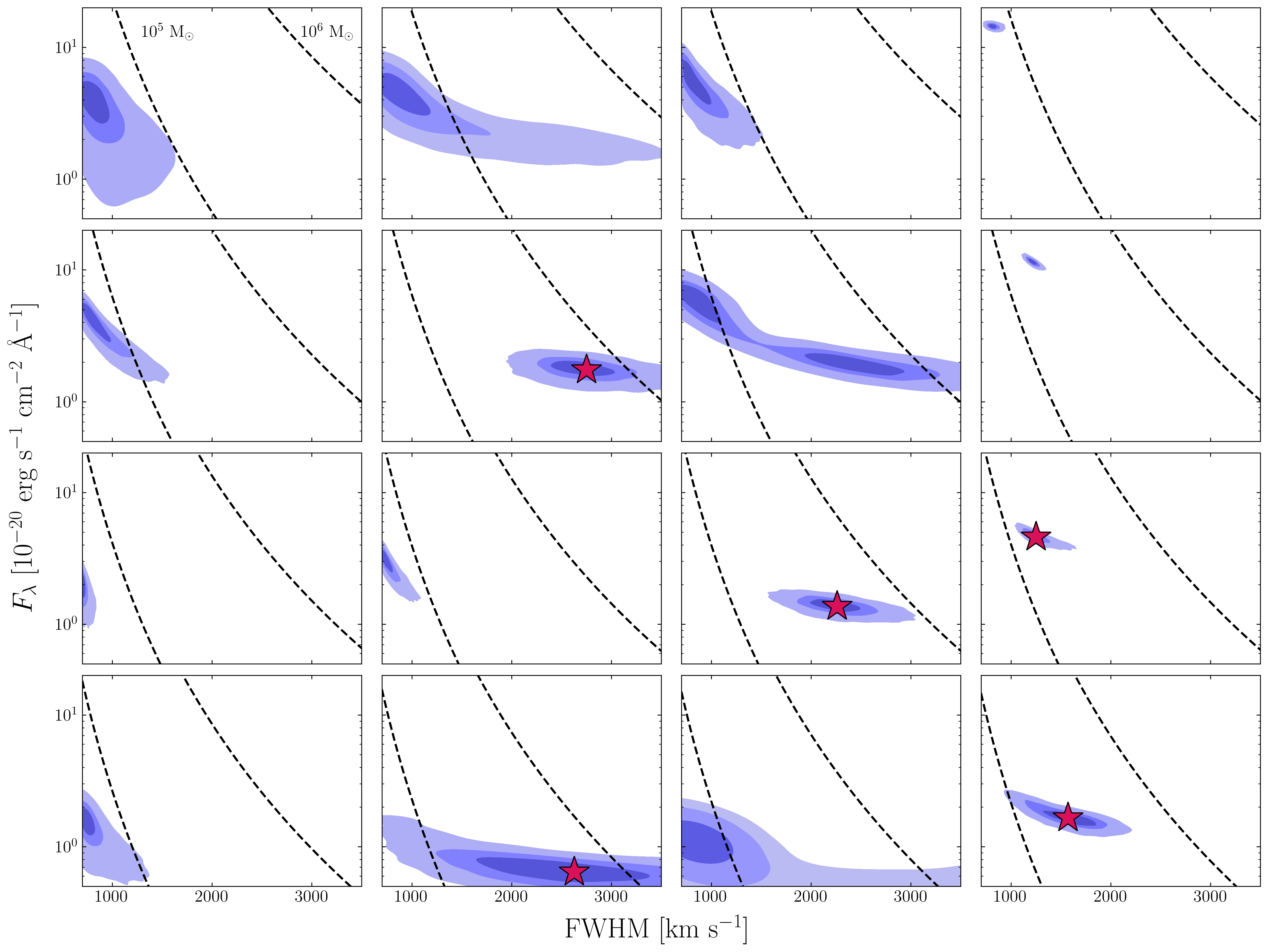}
    \caption{Flux amplitude and FWHM posteriors from our stacked \Ha\ fits. Stacks with detected broad \Ha\ emission have their best fit flux and FWHM measurements shown with the pink star. Full BH mass posteriors are obtained by computing masses with the full flux and FWHM posteriors. BH mass contours of $10^5~\rm{M_{\odot}}$ and $10^6~\rm{M_{\odot}}$ are shown with the dashed black lines. The BH mass prescription we use is discussed in \S \ref{sec: bh masses}.}
    \label{fig: amp_vs_fwhm}
\end{figure*}

\begin{deluxetable*}{ccccccc}
\tablecaption{Median H$\alpha$ Stack Derived Properties}\label{table: measurements}
\tablewidth{20pt}

\tablehead{
  \colhead{Bin} &
  \colhead{$z_{\mathrm{median}}$} &
  \colhead{$\log(\nu L_{\nu,\mathrm{median}})$} &
  \colhead{$\log(L_{\mathrm{H\alpha},\mathrm{broad}})$} &
  \colhead{FWHM} & 
  \colhead{$\log(M_{BH})$} & 
  \colhead{$\log(M_{*})$} 
  \\
  \colhead{} &
  \colhead{} & 
  \multicolumn{2}{c}{[$\mathrm{erg~s^{-1}}$]} & 
  \colhead{[km s$^{-1}$]} &
  \multicolumn{2}{c}{[$M_{\rm{\odot}}$]} 
 }

\startdata
\multicolumn{7}{c}{\text{Broad H$\alpha$ Detections}} \\
6 &
        3.63 &
        42.88 $\pm$ 0.01 &
        $40.17^{+0.06}_{-0.06}$ &
        $2552^{+342}_{-314}$ &
        $5.78^{+0.42}_{-0.42}$ &
        7.84 $\pm$ 0.12
        \\
11 & 
        4.46 & 
        43.20 $\pm$ 0.01 & 
        $40.18^{+0.04}_{-0.04}$ & 
        $1609^{+246}_{-234}$ &
        $5.37^{+0.43}_{-0.42}$ & 
        8.08 $\pm$ 0.14
        \\
12 & 
        4.42 & 
        43.67 $\pm$ 0.10 & 
        $40.55^{+0.02}_{-0.02}$ & 
        $1497^{+90}_{-89}$ &
        $5.48^{+0.40}_{-0.40}$ & 
        8.46 $\pm$ 0.10
        \\
14 & 
        5.62 & 
        43.21 $\pm$ 0.01 & 
        $40.30^{+0.10}_{-0.09}$ & 
        $3510^{+999}_{-877}$ &
        $6.13^{+0.53}_{-0.45}$ & 
        7.88 $\pm$ 0.18
        \\
16 & 
        5.78 & 
        43.82 $\pm$ 0.07 & 
        $40.46^{+0.04}_{-0.04}$ & 
        $1155^{+180}_{-157}$ &
        $5.21^{+0.43}_{-0.41}$ & 
        8.56 $\pm$ 0.03
        \\
\hline
\multicolumn{7}{c}{\text{Non-Detections}} \\
1 & 
        2.13 & 
        42.04 $\pm$ 0.03 & 
        $< 39.83$ & 
        $< 1496$ & 
        $< 4.66$ &
        7.52 $\pm$ 0.10
        \\
2 & 
        2.34 & 
        42.56 $\pm$ 0.01 & 
        $< 39.94$ & 
        $< 2568$ & 
        $< 5.27$ &
        7.99 $\pm$ 0.14
        \\
3 & 
        2.34 & 
        42.98 $\pm$ 0.02 & 
        $< 39.95$ & 
        $< 1495$ & 
        $< 4.84$ &
        8.58 $\pm$ 0.15
        \\
4 & 
        2.47 & 
        43.79 $\pm$ 0.08 & 
        $< 40.45$ & 
        $< 1677$ & 
        $< 5.48$ & 
        8.92 $\pm$ 0.09
        \\
5 & 
        3.68 & 
        42.43 $\pm$ 0.02 & 
        $< 40.25$ & 
        $< 1627$ & 
        $< 5.13$ & 
        7.41 $\pm$ 0.13
        \\
7 & 
        3.69 & 
        43.24 $\pm$ 0.02 & 
        $< 40.49$ & 
        $< 1322$ & 
        $< 5.04$ & 
        8.24 $\pm$ 0.15
        \\
8 & 
        3.63 & 
        43.82 $\pm$ 0.09 & 
        $< 40.77$ & 
        $< 1460$ & 
        $< 5.52$ &
        8.79 $\pm$ 0.14
        \\
9 & 
        4.37 & 
        42.44 $\pm$ 0.02 & 
        $< 40.07$ & 
        $< 994$ & 
        $< 4.66$ &
        7.28 $\pm$ 0.14
        \\
10 & 
        4.46 & 
        42.85 $\pm$ 0.01 & 
        $< 40.26$ & 
        $< 1197$ & 
        $< 4.92$ & 
        7.66 $\pm$ 0.13
        \\
13 & 
        5.28 & 
        42.79 $\pm$ 0.02 & 
        $< 40.25$ & 
        $< 1644$ & 
        $< 4.99$ &
        7.26 $\pm$ 0.26
        \\
15 & 
        5.73 & 
        43.49 $\pm$ 0.01 & 
        $< 41.00$ & 
        $< 14954$ & 
        $< 7.68$ & 
        8.17 $\pm$ 0.08
       \\
\enddata
\tablecomments{We report the $3 \sigma$ upper limit on the $L_{\mathrm{H\alpha},\mathrm{broad}}$, FWHM, and $M_{BH}$ for the stacks with no detected broad \Ha\ emission.}
\end{deluxetable*}

\subsection{Disambiguating BL AGN from Outflows}

Kinematic broadening of the \Ha\ emission line can be caused by galactic-scale outflows from the interstellar medium as well as from the high-velocity gas around an AGN. BL AGN have broad permitted (\Ha, \Hb) lines and narrow forbidden (\OIII) lines \citep[e.g.,][]{VandenBerk2001}, while large-scale outflows would result in broad line components in both permitted and forbidden lines \citep[e.g.,][]{Heckman1990, Veilleux2005, Genzel2014, Veilleux2020}. To confirm the broad \Ha\ lines exhibited in our stacks are caused by AGN rather than outflows, we fit the \OIII $\lambda\lambda4959,5007$ doublet with the same MCMC fitting routine described in \S\ref{sec: Emission Line Fitting}. The \OIII\ line fluxes are fixed to a ratio of 1:2.985 for both the narrow and broad components \citep{Storey2000}. We perform the same $\Delta BIC$ test described in \S\ref{sec: Emission Line Fitting} to determine if a BL component is justified for the \OIII $\lambda\lambda4959,5007$ doublet, indicative of a galactic outflow scenario. Here, $BIC_{narrow} -BIC_{broad+narrow} >~6$ is positive evidence for a broad \OIII\ component and indicates evidence outflows rather than a BL AGN \citep{Liddle2004, fabozzi2014, llerena2023}. We additionally require the detection of the BL component to have a FWHM $> 1000~ \rm{km~s^{-1}}$ and a $3\sigma$ detection of the broad flux. 
Our median \OIII\ stacks are shown in Figure \ref{fig:[OIII] stacks}. 

All (16/16) of our stacks do not show significant evidence for a broad \OIII\ component, with similar kinematics to broad \Ha, as they fail all three of our detection requirements ($\rm{FWHM > 1000~km~s^{-1}}$, $3\sigma$ detection of $f_{broad}$, and $\Delta BIC > 6$). We find that the distribution of narrow FWHMs for \Ha\ and \OIII\ are consistent across our bins. This indicates for no outflow-driven kinematics present in our stacks and suggest the broad \Ha\ component is tracing Doppler broadening from an AGN.

\vspace{0.1cm}

\section{Results}\label{sec: results}

In this study, we create 16 median-stacks of \Ha\ and \OIII, binned by redshift and continuum luminosity, from spectroscopy in the JWST extragalactic deep fields. These stacks have median redshifts ranging from $z = 2.13 - 5.78$ and continuum luminosities $\log(\nu L_{\nu}) = 42.04 - 43.82~ \rm{erg~s^{-1}}$. In 5 of these stacks, we find evidence for a broad \Ha\ emission component, with no corresponding broad \OIII\ component.
The detection of a \Ha\ BL component in our stacks provides strong evidence for the presence of weak AGN in many high-redshift galaxies that are undetected in individual spectra.

\subsection{BH masses}\label{sec: bh masses}

\begin{figure*}[]
    \centering
    \includegraphics[width=.9\linewidth]{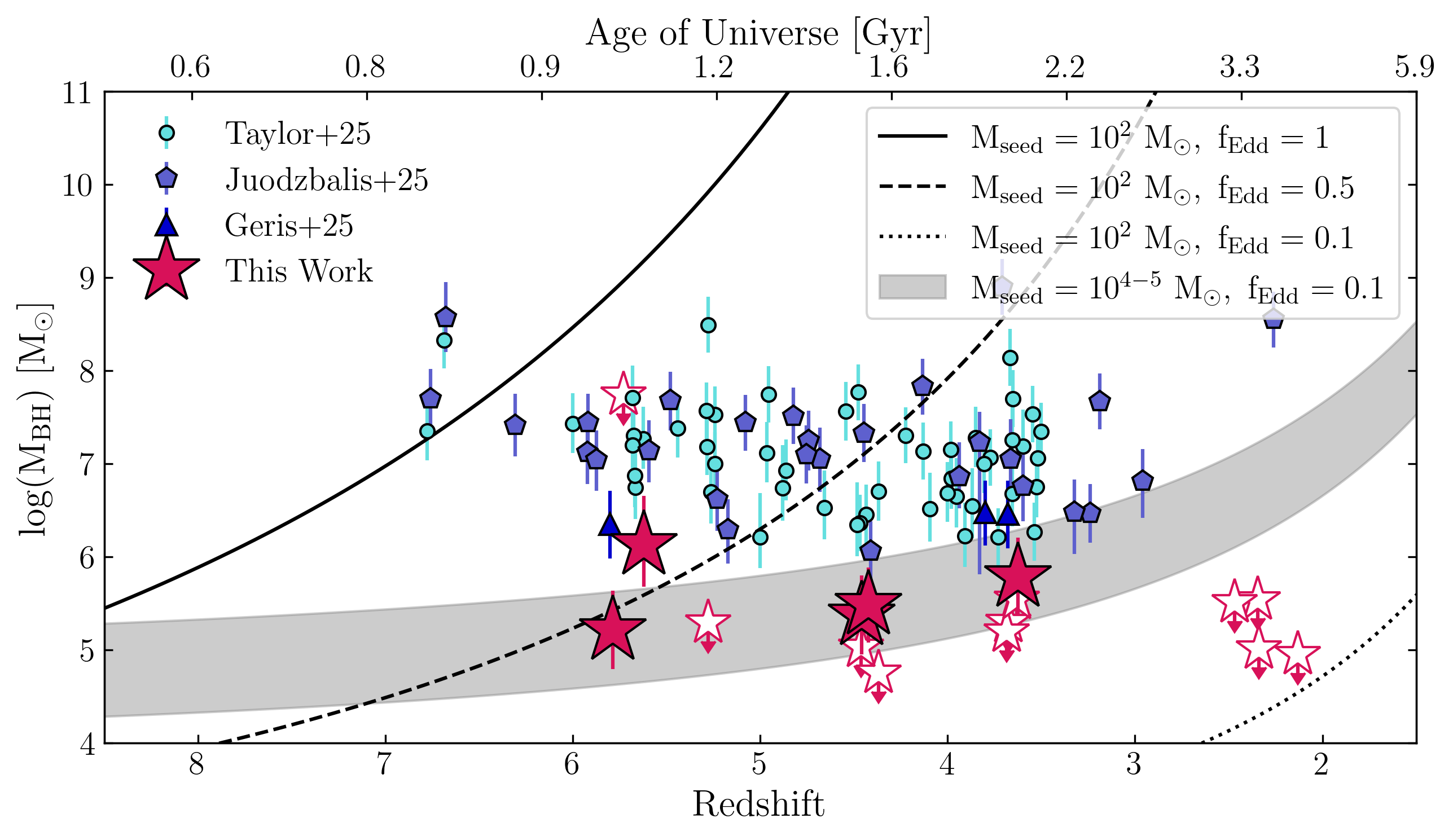}
    \caption{BH mass as a function of redshift. The broad \Ha\ line detections from our stacks are shown with the red stars, and $3\sigma$ upper-limits for the non-detections are shown by the open stars. Two large JWST BL AGN surveys are shown with the light blue circles \citep{Taylor2024} and the purple pentagons \citep{Juodzbalis2025}. The highest and second-highest \OIII\ luminosity stacks from \cite{Geris2025} are shown with the cyan squares. We show the growth of a heavy $10^{4-5}M_{\odot}$ BH seed with  $f_{Edd} = 0.1$ (gray shading), which is the typical Eddington ratio at lower redshifts \citep{Ananna2022}. We show three growth paths of a Population III stellar remnant ($10^2 M_{\odot}$) with $f_{Edd} = 1$ (solid black line), $f_{Edd} = 0.5$ (dashed black line), and $f_{Edd} = 0.1$ (dotted black line). In our models, the stellar remnant formed at $z = 30$, with accretion beginning after a 100 Myr delay.}
    \label{fig:mbh_redshift}
\end{figure*}

We measure BH masses for our stacks using the following prescription from \cite{Reines2013}:

\begin{multline}
    M_{\rm{BH}} = 10^{6.57} \times \left(\frac{L_{\mathrm{H} \alpha, \rm{broad}}}{10^{42} \mathrm{~ergs} \mathrm{~s}^{-1}}\right)^{0.47} \\ 
    \times \left(\frac{\mathrm{FWHM}_{\mathrm{H} \alpha, \rm{broad}}}{10^{3} \mathrm{~km} \mathrm{~s}^{-1}}\right)^{2.06} M_{\odot}
\end{multline}
where $L_{H\rm{\alpha,\rm{broad}}}$ is the luminosity of the broad \Ha\ component and $\rm{FWHM}_{\rm{H\alpha, broad}}$ is the FWHM of the \Ha\ broad component. To measure the \Ha\ luminosity for each stack, we use the median redshift of the stack and re-normalize the stack by the median \Ha\ line flux.

The BH masses found from broad \Ha\ emission line measurements rely on single-epoch virial relations derived from the local Universe \citep{Greene2005}. These measurements depend on the assumptions that the gas around the BH is indeed virialized and that the BL radius ($R_{\rm BLR}$) and luminosity relation hold at high redshifts. Due to the symmetric nature of the broad emission lines found in this sample, we find that the kinematics are consistent with the basic virial assumption underlying the BH mass equation. Additionally, \cite{Ji2025} finds for a $z \simeq 7$ ``little red dot" (LRD: \cite{Mathee2023}) that the $R_{\rm BLR}$-luminosity relation is consistent with local reverberation mapping and \cite{Juodzbalis2025b} reports a direct BH mass measurement at $z = 7$ that is in agreement with the single-epoch virial estimate. We therefore use the local virial relations to compute canonical BH masses for our sample, but acknowledge the uncertainties associated with their use at high redshifts.

To make our BH measurements and to best propagate the uncertainties in our \Ha\ line fitting, described in \S \ref{sec: Emission Line Fitting}, we compute BH masses from the full broad \Ha\ flux and FWHM posterior distributions. In doing so, we have a full posterior distribution of BH masses for each stack. We additionally add a 0.4~dex uncertainty in quadrature to our BH mass errors to account for systematic scatter in the BH mass equation \citep{Vestergaard2006}. For our stacks with no broad \Ha, we derive the $3\sigma$ BH mass upper limit from the 99.7 percentile of the posterior distribution. The posterior distribution of our flux and FWHM values for our stacks are shown in Figure \ref{fig: amp_vs_fwhm}. The stacks with a detected broad \Ha\ emission line have their best-fit flux and FWHM shown with a pink star. BH mass contours of $M_{BH} = 10^5$ and $10^6~\rm{M_{\odot}}$ are shown.

We measure median BH masses from the 5 broad \Ha\ detections in our stacks of $\log(M_{BH}) = 5.27 - 6.13~ \rm{M_{\odot}}$. These stacks are pushing the detectable BH mass regime to $\lesssim 10^6~\rm{M_{\odot}}$, which due to the sensitivity of JWST is too faint to detect in the individual NIRSpec spectra (without magnification from gravitational lensing \citep{Fei2025}) of current JWST surveys.

\begin{figure*}
    \centering
    \includegraphics[width=.6\linewidth]{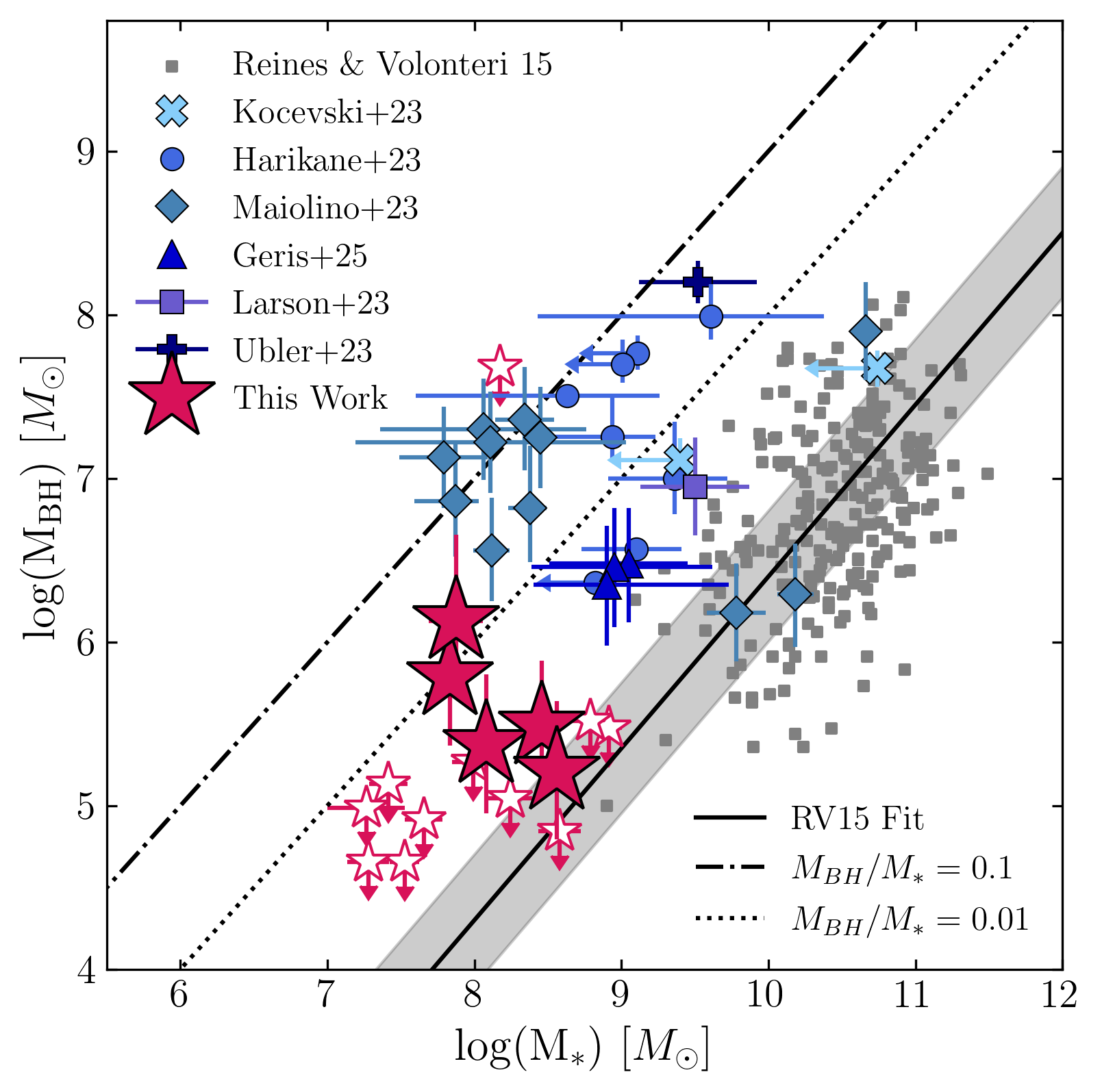}
    \caption{BH mass as a function of stellar mass. Stacks with detected broad \Ha\ emission are shown with the red filled in stars, and $3\sigma$ upper limits of BH mass are shown with the red open stars. The gray squares show local AGN from \cite{ReinesVolonteri2015}. Other JWST detected BLAGN are shown: \cite{Larson2023} (purple square), \cite{Ubler2023} (plus), \cite{Harikane2023} (circles), \cite{Maiolino2023} (diamonds), and \cite{Kocevski2023} (crosses). Another stacking analysis done by \cite{Geris2025} in the JADES deep fields is shown by the blue triangles. The $M_{BH}-M_*$ relation derived by \cite{ReinesVolonteri2015} (low-$z$ relation) is shown by the solid black line. We also show $M_{BH}/M_{*} = 0.1$ (dashed-dotted line) and $M_{BH}/M_{*} = 0.01$ (dashed line).}
    \label{fig:mbh_mstar}
\end{figure*}

\section{Discussion}\label{sec: discussion}
In this work, we present the median-stacked spectra of $\sim2000$ galaxies ($\sim125$ galaxies per stack) in the JWST extragalactic deep fields to search for the broad \Ha\ signature of AGN emission within the broader galaxy population (without BL AGN detections in their individual spectra). In $31\%$ (5/16) of our stacks, we find evidence for broad \Ha\ emission-line components that indicate a population of lower mass ($<10^6~\rm{M_{\odot}}$) BHs.
We also derive BH mass upper limits for our stacks with no detected broad \Ha\ component. Through this stacking approach, we are able to probe the AGN occupation of ``normal" galaxies that is not represented by the luminous individual AGN detections.  

\subsection{BH Seeding}

The most luminous AGN detected with JWST at high redshifts provide interesting constraints for BH seeding models \citep[e.g.,][]{Pacucci2022}. Most can be described by either a massive seed growing at the Eddington rate or a stellar remnant seed undergoing super-Eddington accretion \citep[e.g.,][]{Larson2023, Maiolino2024a, Taylor2025}. Heavy seed formation from direct-collapse black holes (DCBHs) requires unusual halos with pristine low-metallicity gas and low angular momentum \citep[e.g.,][]{Woods2019}, and could be too rare to describe all the JWST AGN being observed at these epochs \citep{Bhowmick2024}. Stellar dynamical processes in nuclear star clusters could also produce a heavy black hole seed \citep[e.g,][]{Sanders1970, Begelman1978}, which observations suggest is more common at high redshift \citep{Mowla2024, Adamo2024, Fujimoto2025}. A nuclear star cluster BH seed could explain the observed masses and properties of the LRD population observed in the high-redshift Universe \citep{Pacucci2025}.

Figure \ref{fig:mbh_redshift} shows our BH mass measurements as a function of redshift, along with two large JWST BLAGN surveys of individually-detected BL AGN \citep{Taylor2024, Juodzbalis2025} in the $ 2 < z < 7$ regime, and another stacking analysis in the JADES deep fields performed by \cite{Geris2025}. Our AGN detections are shown with the red filled-in stars, and the BH mass $3\sigma$ upper limits for our non-detections are shown with the open stars. We also plot simple models of BH growth for both light ($\sim 10^2~\rm{M_{\odot}}$) and heavy ($> 10^4 ~\rm{M_{\odot}}$) BH seeds. The solid, dashed, and dotted line represents a light seed ($M_{seed} = 10^2~\rm{M_\odot}$) with $f_{Edd} = 1,~0.5,~\rm{and}~0.1$, respectively. The stellar remnant in the light seed models forms at $z=30$ and starts accreting after a 100 Myr delay. Additionally, the gray shaded region represents a $10^{4-5}~\rm{M_{\odot}}$ heavy seed with $f_{Edd} = 0.1$ \citep{Ananna2022} that formed at $z=15$.

 We find that the median BH masses implied by our BL AGN detections could be formed by light stellar remnant seed ($10^2~\rm{M_{\odot}}$) growth models with $f_{Edd} = 0.1 - 0.5$. This suggests that the median galaxy at high redshifts does not require rare or exotic seeding mechanisms to explain the mass of its BH, but can be described by a stellar remnant seed undergoing Eddington-limited accretion. The BL AGN detection in the highest-redshift ($z=5.78$) bin places the strongest constraint on BH seeding modelings, but is still consistent with formation from a light seed growing with $f_{\rm Edd} \simeq 50\%$. The BL AGN detections are also consistent with the formation of a heavy seed that grows at a lower accretion rate (gray shaded region).
 
 %Additionally, we find that our highest redshift ($z = 5.78$) detection can be described by a light seed with moderate-Eddington accretion ($50\%$) or is well-fit by a heavy seed model.

 \begin{figure*}
    \centering
    \includegraphics[width=.8\linewidth]{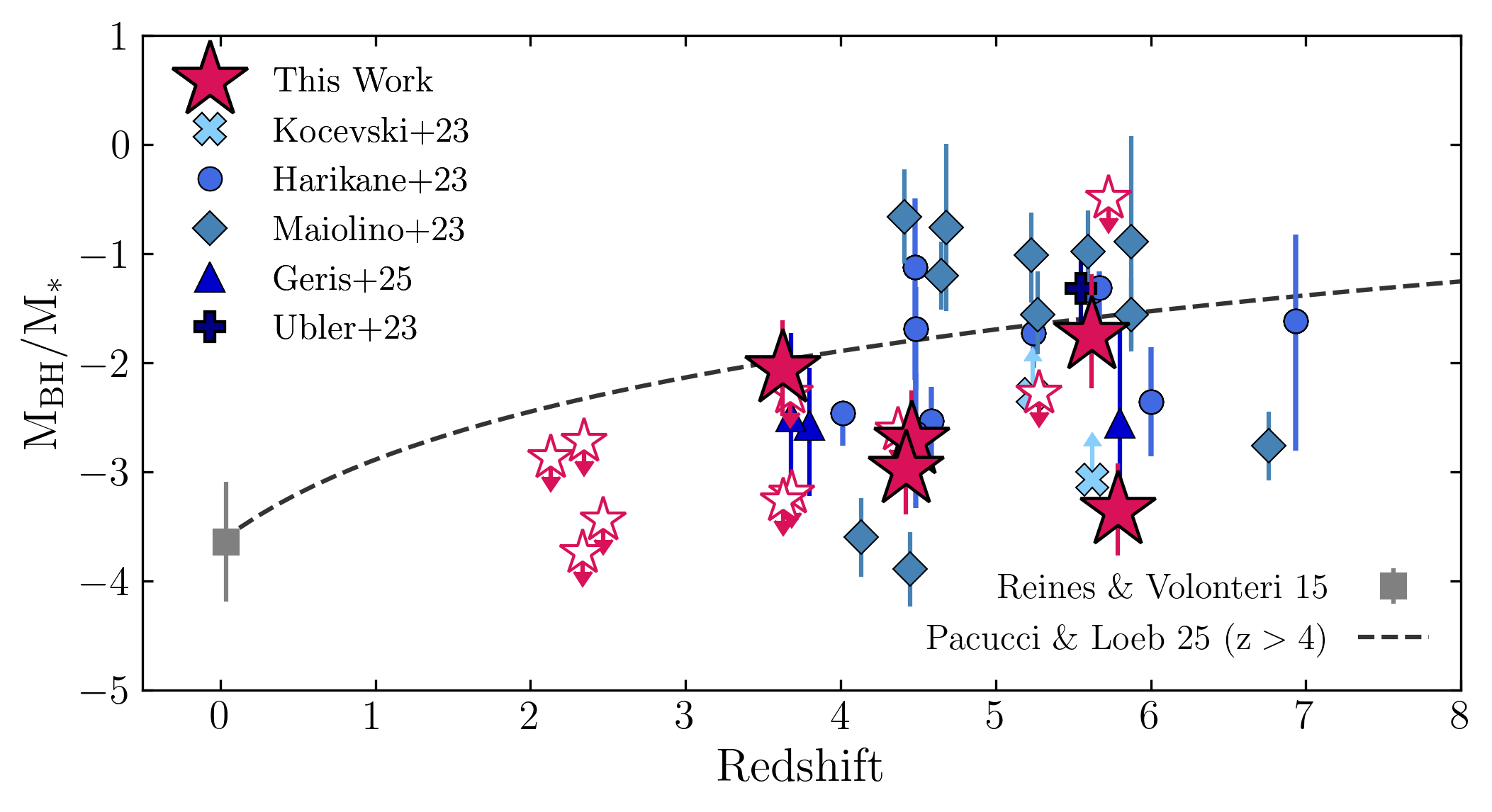}
    \caption{$\rm{M_{BH}/M_{*}}$ as a function of redshift. Stacks with detected broad \Ha\ emission are shown with the red filled in stars, and $3\sigma$ upper limits of BH mass are shown with the red open stars. Other JWST detected BLAGN are shown: \cite{Ubler2023} (plus), \cite{Harikane2023} (circles), \cite{Maiolino2023} (diamonds), \cite{Kocevski2023} (crosses). Another stacking analysis done by \cite{Geris2025} in the JADES deep fields is shown by the triangles. We show the mean of the local AGN population \citep{ReinesVolonteri2015} with the filled in gray square. We also show a high-$z$ ($z > 4$) scaling relation from \citep{PacucciLoeb2024} with the dashed black line.}
    \label{fig:ratio_vs_redshift}
\end{figure*}

\subsection{AGN Activity as a Function of Redshift}
We detect a BL component in 5 of the \Ha\ stacks with a median redshift $z > 3.68$, and none in the redshift range $z = 1.58 - 3.33$. This is indicative of finding a hidden AGN population in luminous high-redshift galaxies but not in the lower redshift galaxies.

This trend suggests that these lower redshift BHs could have lower accretion rates. If broad emission lines are present only in the most rapidly accreting AGN \citep{Trump2011, Elitzur2014}, then the absence of a broad \Ha\ detection could indicate that these AGN are accreting at relatively low Eddington ratios. \cite{ShenKelly2012} similarly find evidence for higher AGN Eddington ratios (at fixed BH mass) at higher redshifts among SDSS quasars.

The lack of AGN activity seen in the lower-redshift stacks could also indicate the rise of galaxy-scale obscuration \citep{Whitaker2017} in more recent epochs. This obscuration could be different than the exotic obscuration mechanisms seen in LRDs, like a dense gas cocoon surrounding a luminous AGN \citep{Inayoshi2025, Naidu2025, deGraaff2025, Taylor2025}, but would still obscure the AGN even if the obscuring material is on galactic-scales. LRDs have well-detected \Ha\ emission with weak \Hb\ and UV lines, consistent with an obscuring medium that is attenuating the bluer part of the spectrum \citep{Brooks2025}, which is different than the total broad \Ha\ obscuration we observe in our lower redshift stacks.

%%    Additionally, the detection of lower mass BHs ($<10^6~\rm{M_{\odot}}$) in our higher redshift stacks could potentially be the ancestors of SDSS quasars \citep{Shen2011}. If these AGN continue to grow at moderate efficiency, they ...

\vspace{0.2cm}
\subsection{The $M_{BH}-M_*$ Relation}

JWST observations of high-redshift AGN have shown a departure from the tight low-z BH mass-stellar mass relation. Most individually-detected BL AGN are dramatically over-massive compared to the local $M_{BH}-M_*$ \citep[e.g.,][]{Maiolino2023, Kocevski2023,Pacucci2023, Harikane2023,Larson2023, Durodola2025, Chen2025, Jones2025, Cohn2025}. These observations are subject to significant selection bias, since only the most luminous AGN can be detected in current JWST surveys and these represent only the rare tail of the larger AGN population \citep{Li2025, Sun2024}. Measuring the global $M_{BH}-M_*$ relation requires a more sophisticated approach that probes AGN content across the broader galaxy population: i.e. this work.

%These observations are subjected to selection bias, which contributes to the proposed evolution of the $M_{BH}-M_*$ relation at high redshift. The true evolution of the $M_{BH}-M_*$ relation remains highly debated.

Figure \ref{fig:mbh_mstar} shows the BH mass-stellar mass relation for our 16 stacks and Figure \ref{fig:ratio_vs_redshift} shows $M_{BH}/M_*$ as a function of redshift. Each $M_{BH}$ and $M_*$ measurement represents a median value (from median-stacked spectra and SEDs) for $\sim125$ galaxies in bins of redshift and continuum luminosity. We show our detections with the filled red stars and our BH mass $3\sigma$ upper limits with open red stars. We additionally plot individually-detected luminous BL AGN from other JWST surveys \citep{Kocevski2023, Harikane2023, Maiolino2023, Larson2023, Ubler2023}, another \Ha\ stacking analysis \citep{Geris2025}, and the low-$z$ AGN population from \citet{ReinesVolonteri2015}.

The BH masses derived from the \Ha\ median-stacked spectra probe into a lower black hole mass regime ($< 10^6~\rm{M_{\odot}}$) than the individual BL AGN detections alone. These BHs also lie in host galaxies with a median stellar mass of $10^{7.8} - 10^{8.6}~\rm{M_{\odot}}$, which are estimated from the median stacked SEDs that assume no AGN contamination, described in \S \ref{sec: stellar masses}. Two of our stacks with detected broad \Ha\ lie within the scatter of the local relation, while the other three stacks with AGN detections lie $\sim1$ dex above the local relation. This suggests that the stacked spectra are less ``over-massive" (\textit{at most} 10 times over-massive) than previously discovered JWST AGN, which lie 2-3~dex above the local relation. %Additionally, one of our BH mass upper limits derived from the non-detection stack falls on the local relation. 
The broad \Ha\ limits similarly imply BHs that are over-massive by a factor of $\lesssim 10$ and/or consistent with the low-$z$ $M_{BH}-M_*$ relation. Our results indicate that individual detections of rare and luminous AGN are more likely to sample the tail end of the $M_{BH}-M_*$ distribution at high redshifts, while stacking on the non-detected AGN population overcomes this selection bias and reveals that the median high-$z$ galaxy does not host a dramatically over-massive black hole.

Our results are also consistent with \cite{Ren2025}, which identified 7 AGN candidates through JWST/NIRSpec IFU that have BH masses of $10^6 - 10^{7.5}~\rm{M_\odot}$ and stellar masses of $10^{9.5} - 10^{10.5}~\rm{M_{\odot}}$. \cite{Ren2025} circumvent selection bias through spatially-resolved NIRSpec IFU data, which allows them to minimize host-galaxy contamination in their BL detection methodology. \cite{Jones2025} find that non-LRD AGN are consistent with the local scaling relation and LRD selected BL AGN tend to be more over-massive, which suggests that our population probed by stacking is more similar to the non-LRD population. In addition to \cite{Ren2025} and \cite{Jones2025}, our work is also consistent with \cite{Li2025} and \cite{Sun2024}. In this interpretation, the $M_{BH}-M_*$ relation may have increased scatter but not increased normalization at high-$z$ \citep{Peng2007,Hirschmann2010,Jahnke2011}. Our results imply that there is a larger fraction of galaxies with over-massive BHs, but median galaxies have more lower-mass BHs that are consistent or perhaps modestly over-massive with respect to the local relation. In this increased-scatter scenario for $M_{BH}-M_*$, there is also likely an unseen population of galaxies with under-massive BHs. 

% The lower mass range of BHs probed by this study, \cite{Ren2025} and \cite{Geris2025}, are predicted by simulations.

%   The complete census of the detected galaxy population indicates a quite different story from the overwhelmingly massive BHs discovered through luminous AGN. The extreme massive AGN observed at high redshift might be indicative of a larger intrinsic scatter at these epochs. The hierarchical merging of these galaxies, even those on the extreme end of the $M_{BH}-M_*$ distribution, will eventually form a tighter relationship, as seen from  local AGN \citep{Jahnke2011}.

\section{Conclusions}

The large population of luminous AGN discovered in the early Universe is one of the biggest surprises and scientific advancements of JWST \citep[e.g.,][]{Larson2023,Kocevski2023,Ubler2023,Harikane2023, Maiolino2023, Taylor2024, Juodzbalis2025}. This work investigates the AGN occupation of ``normal" galaxies by selecting on non-broad \Ha\ emission and finds that the median galaxy lies closer to the local $M_{BH}-M_*$ relation. Our main findings are summarized below:
\begin{itemize}
    \item We find evidence for a broad \Ha\ emission component in $31\%$ (5/16) of our stacks and no evidence for a broad \OIII\ component, suggesting the presence of weak AGN in many high-redshift galaxies.
    \item We measure median BH masses from the broad \Ha\ detections of $10^{5.21} - 10^{6.13}~ \rm{M_{\odot}}$. From the stacked SEDs, we measure median stellar masses of $10^{7.8} - 10^{8.6}~\rm{M_{\odot}}$.
    \item The BH masses derived from our stacked broad \Ha\ detections are well described by light stellar remnant seeds undergoing Eddington-limited accretion. Our highest redshift ($z =5.78$) is also consistent with a light seed growing with $f_{Edd} \simeq 50\%$.
    \item The BH and stellar masses of our stacked detections suggest that the median galaxy hosts a BH that is \textit{at most} a factor of 10 times over-massive, while individually detected AGN can be up to 100 times over-massive when compared to the local relation. This result indicates that individual detections of AGN are much more likely to sample the high-$z$ tail-end of the $M_{BH}-M_*$ relation.
    
\end{itemize}

This work emphasizes the importance of selection bias when studying the evolution of the $M_{BH}-M_*$ relation across cosmic time. Deeper and higher resolution observations of early Universe BHs will provide critical information to further interpret the BH-galaxy connection at these epochs. 

\section{Acknowledgments}
(Some of) The data products presented herein were retrieved from the Dawn JWST Archive (DJA). DJA is an initiative of the Cosmic Dawn Center (DAWN), which is funded by the Danish National Research Foundation under grant DNRF140. We thank the JADES, GLASS, PRIMER, UNCOVER, and RUBIES team for their effort designing and executing their programs and for making the data publicly available.

MB acknowledges support from a NSF Graduate Research Fellowship award number 2136520. MB, JRT, and KD acknowledge support from NSF grant CAREER-1945546 and NASA grants JWST-GO-06368 and JWST-GO-05718. RA acknowledges support of grant PID2023-147386NB-I00 funded by MICIU/AEI/10.13039/501100011033 and by ERDF/EU, and the Severo Ochoa grant CEX2021-001131-S to the IAA-CSIC.

\facility{\textit{JWST}}
\facility{\textit{HST}}
\software{\texttt{astropy}: \cite{Astropy2013, Astropy2018, Astropy2022}, \texttt{scipy}: \cite{scipy2020}, \texttt{emcee}: \cite{emcee}, \texttt{numpy}: \cite{numpy}, \texttt{matplotlib}: \cite{matplotlib}}

% \clearpage
 \begin{appendix} \label{appendix}

\begin{figure*}\label{fig:stellar mass SEDs}
       \centering
       \includegraphics[width=\linewidth]{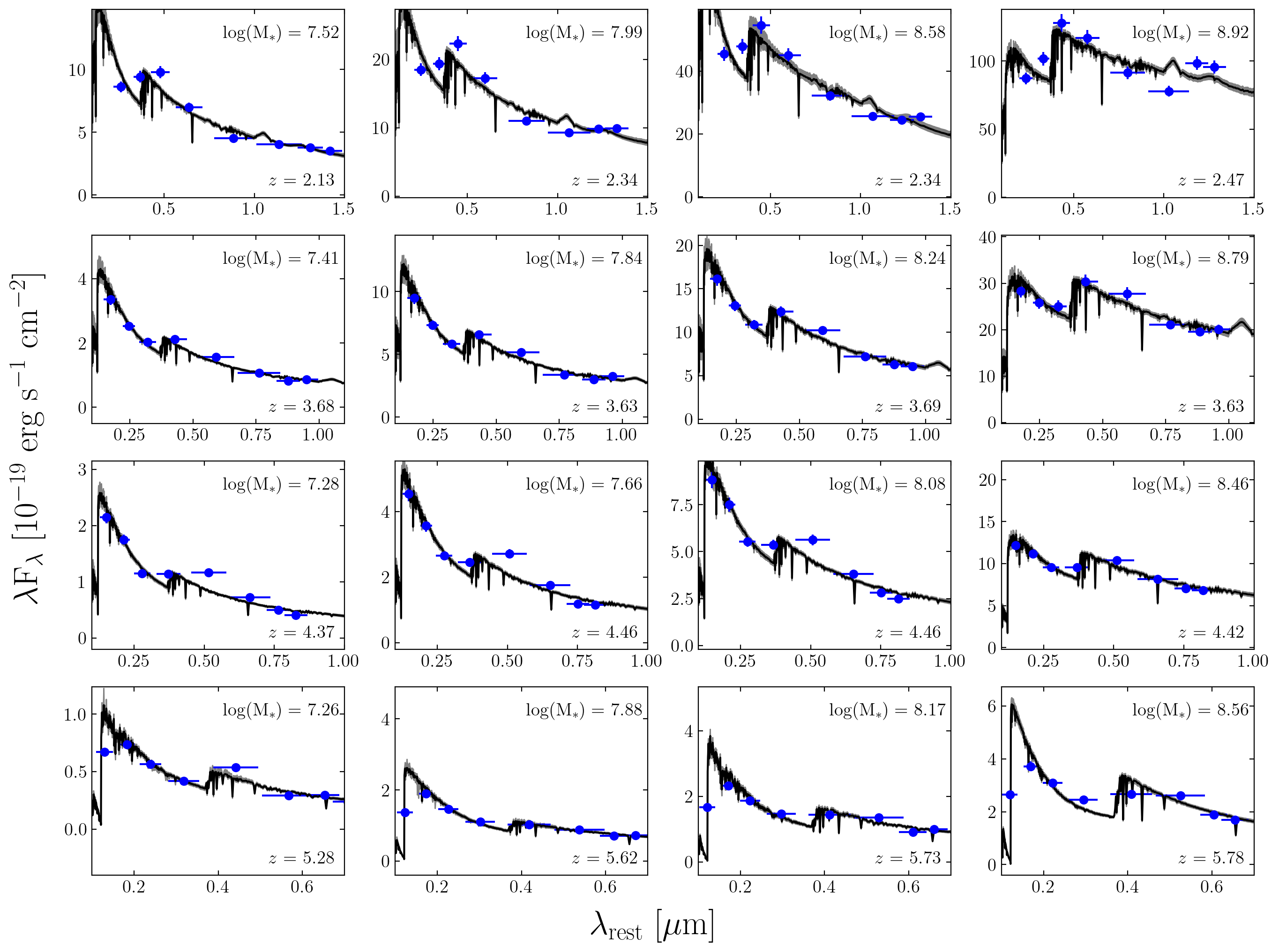}
        \caption{$\lambda F_{\lambda}$ vs $\lambda_{\rm{rest}}$ for our 16 NIRCam+HST stacks. Blue points are the median-stacked fluxes and the best-fit SED is shown in black. Our stacking process is described in \S \ref{sec: stacked spectra} and our stellar mass measurements are described in \S \ref{sec: stellar masses}. The horizontal errors shown represent the filter widths. }
    \end{figure*}

 Figure \ref{fig:stellar mass SEDs} shows the median-stacked SEDs and best-fit \texttt{bagpipes} models for our 16 bins. Our stacking procedure is described in \S \ref{sec: stacked spectra} and our SED fitting is described in \S \ref{sec: stellar masses}.
 
\end{appendix}

\bibliographystyle{aasjournal}
\bibliography{citations}

\end{document}